# ECONOPHYSICS IN BELGIUM. THE FIRST (?) 15 YEARS

M. AUSLOOS*†

*This reviews the econophysics activities in Belgium from my admittedly biased point of view. Unknown historical notes or facts are presented for the first time explaining the aims, whence evolution of the research papers and friendly connexions with colleagues. Comments on endeavours are also provided. The lack of official, academic and private support is outlined.*

***Preliminary note:*** When thanking the editors for their invitation, I sent them a warning. It was not obvious to me whether econophysics in Belgium was really a good title. I am not much aware about econophysics research activity in Belgium, except for the one in Liège. This might be attributed to an overemphasised ego of mine, or to my lack of interest for the work of nearest neighbours. Not believing so, I am afraid that I have to summarise and comment on my own activities in the field for the last 15 years or so.

## Introduction

In 1987, a giant intellectual tsunami shocked condensed matter physics, following an unexpected scientific earthquake[1] the previous year: superconductivity could be found above liquid nitrogen temperature[2]. In Liège, in the high precision electric measurement and instrumentation laboratory (MIEL) of the Faculty of Applied Sciences, one had a set-up and the aim to measure finely electrical resistivity, among other electrical transport properties. Only magnetic samples had up to then be investigated. Samples were coming from outside the University of Liège (ULG) chemical laboratories. However there was a glass crystal growing group in the chemistry laboratory of the Faculty of Sciences. Even though the professor of chemistry had never been one of my supporters in my scientific aims and activities, I decided to talk to him about growing some $Y Ba_2Cu_3O_{7-x}$ ceramics (YBCO). He agreed; he had some collaboration with B. Raveau in Caen, and had some so called 211 green powder directly available. There is no space to explain how things successfully developed and our findings of critical exponents in a nevertheless polycrystalline YBCO sample[3–6]. High Critical Temperature Superconductors (HCTS) research activities went on through a joint endeavour of the two laboratories, yet depending of different deans, and with *a priori* different academic specificities. One technique which was developed was to grow such HCTS but replacing *Y* by a magnetic rare earth ion, e.g. Dy. Since some electro-magnet was available at the MIEL, ceramics crystal growth under a magnetic field was attempted[7–10].

With some success microscopic observations led to unexpected crystallographic features[11]. It was not easy to explain them, except through intuition. No model was available. No crystal growth equations were satisfactory for explaining the features. Moreover it takes a long time to grow such ceramics which after the initial growth at high temperature, under strict conditions, must be slowly and well oxygenated. The process takes a very long time. After that, chemical and structural characteristics, electrical and magnetic measurements had to be made[12]. Without necessarily finding something drastically different. A conceptual imagination of some process modification was thus harduous to implement without being sure of some success. Another theoretical approach had to be found.

I had already proposed to the Faculty of Sciences responsible for teaching, in the 80's, to develop a "free to any audience" series of lectures on *Fractals* in 1986. One of the lectures was on kinetic growth models[13], like the Eden Model[14]. From my graduate years, I had easily learned that if one is not too intelligent and does not belong to an internationally famous physics group, one has to add to what was previously imagined by others. There are basically two ways to do so. One is to introduce some extra degree of freedom in some model; the next one which somewhat goes with it is to introduce an external field which couples to the additional degree of freedom. This had already been done[15] in the theoretical group which I had attempted to set-up in Liège, when magneto-fluids appeared to be an interesting field of research.

* 7 rue des Chartreux, B-4122 Plainevaux, Belgium
† previously at GRAPES@SUPRATECS, Université de Liège, B5a Sart-Tilman, B-4000; Liege, Euroland; e-mail address: marcel.ausloos@ulg.ac.be

It appeared that one could imagine a Magnetic Eden Model (MEM)[16–18] such that the growth element had a magnetic degree of freedom, a spin, which could thus be coupled to a magnetic field. This was easily simulated numerically[10,19–21]. Wonderful agreement was found between the crystallographic features of HCTS Dy-based ceramics and the numerical results. However in order to validate the MEM faster, it would have been fine not to wait for the results of physical measurements. What are basically needed is a time series indicating some growth evolution. Once in a queue at a bank the word growth came to mind. Banks were places where growth was ... a must; should data be available?

### *Money, Money, Money, … a well known song*

**Foreign currency exchanges :** I tried to convince my banker that if I could get some data, I would analyse them with respect to my kinetic growth model *mutatis mutandis*. I was told that there would be some problem because of bank confidentiality status, and client privacy status. Thus, no, no data would be available. However much data is available on e.g. Yahoo. Whence the first data analyzed *ca.* 1996, was about coherent and random sequences in daily closing USD/DEM exchange rate fluctuations[22], within the moving window Detrended Fluctuation Analysis (DFA) [23]. It was shown that the evolution of the DFA exponent away or toward a half-odd-integer value could be tied to speculation after major economic meetings or drastic events. Thereafter some discussion took place concerning predicting fluctuations in foreign exchange currency rates. A blind game was played with the then major bank in Belgium, receiving exchange rate values by fax in the evening from J. Pirard. After some DFA, the most reasonable prediction was sent by fax to the exchange room the next morning. It had been agreed with the bank management, e.g. P. Praet, later to become Governor of the National Bank, that after three months the bank results and ours would be compared, but at the June 97 meeting in Brussels the bankers refused to compare results, without giving any reason why.

**Crashes :** In the mean time having been aware of the paper about the ionic distribution in water before the Kobe earthquake and the somewhat unexpected logperiodic oscillations (LPO) in such "signals"[24], a monitoring of the S&P500 and DJIA was made as early as April 1997. The Monday Oct. 19, 1987 stock market crash was still vividly present: on that day, the Dow Jones Industrial Average (DJIA) lost 21.6 %. The 1997 financial indices evolution was very similar to those of 1987 and a LPO signature was obvious. The description of the 1987 stock market crash had been already proposed in two independent works[25,26]. They indicated that the economic index $y(t)$ follows a power law with a complex exponent, i.e. like

$$y(t) = A + B\left(\frac{t_c - t}{t_c}\right)^{-m} \left[1 + C\left(w \ln\left(\frac{t_c - t}{t_c}\right) + \phi\right)\right] (t < t_c). \quad (1)$$

where $t_c$ is the crash-time or rupture point, - the other symbols being parameters. A divergence occurs at $t = t_c$ if the exponent $m > 0$; moreover the oscillations condense on one day which is the rupture point $t_c$. This law is similar to that of critical points at so-called second order phase transitions[27], but generalizes the scaleless situation when discrete scale invariance exists[28].

Fits using Eq.(1) were performed on the S&P500 data[26,29] for the period preceeding the 1987 October crash. The results are not robust against small perturbations due to the instability of the nonlinear seven parameter fit. For ease the "universal" exponent $m$ was imposed to be zero, i.e. The divergence of an index $y$ for $t$ close to $t_c$ should rather read

$$y(t) = A + B \ln\left(\frac{t_c - t}{t_c}\right)\left[1 + C\left(w \ln\left(\frac{t_c - t}{t_c}\right) + \phi\right)\right] (t < t_c). \quad (2)$$

The argument is based on physics: a logarithmic behaviour is known in physics as characterizing the specific heat ("four point correlation function") of the magnetic Ising model[30]. Another type of essential singularity in physics is that occurring at the Kosterlitz-Thouless (KT) phase transition[31] for dislocation mediated melting. It represents a transformation from a disordered vortex fluid state with equal number of vortices with opposite "vorticity" to an ordered molecular-like state with molecules composed from a pair of vortices with different polarities across the KT transition. The logarithmic behavior can also be observed in the temperature derivative of the electrical resistivity of magnetic systems at the paramagnetic-ferromagnetic phase transition[32]. The behaviour is thus generally specific to systems with a low order spatial dimensionality of the so-called "order parameter".

In order to test the validity of Eq.(2) in the vicinity of crashes, the 6-parameter function was "separated" into monitoring the divergence itself and the oscillation convergence on the other hand. The rupture point $t_c$ had to come from an agreement between both fits , within some error bars. In so doing in fact, the long range and short range fluctuation scales are examined on the same footing.

Three 1987 indices were first tested. The, so called predicted, rupture points were located very close and just after the real date of the October 1987 crash[33]. Next, for the period January 1990 - September 1997 the stock market conditions astonishingly looked like the pre-crash period of 1987. The rupture points were estimated for November - December 1997, in the domain of expectations.

A paper was submitted to a major scientific journal with one of the highest scientific impact factors, predicting a crash in the fall of 97. The paper was refused with the usual classical statement from the editor when turning back such a work. Being more and more convinced of the crash occurrence[a], I turned toward an ex-physics student who had become a journalist for one of the weekly leading business journals in Belgium. In August a sort of interview was set-up in a small room of the new main auditorium building on the Sart-Tilman campus, and the journal came up with a title asking whether the stock market could be put into equations[34] and indicating the crash to occur by the end of October, with comments in[35].

Subsequent monitoring allowed the sending of a warning fax to the journal editor on Saturday Oct. 25 in the morning; the crash occurred on Monday Oct. 27. The journal coming up the following week had an editorial entitled "Nous l'avions annoncé" ("We predicted it"). It was not immediately obvious who was "we"[36], with comments in[37]. Next, to publish papers how on the subject in scientific journals had us to accept a stringently imposed (by the referee and the editor) modification of the title from *"How the crash of Oct. 27, 1997 was predicted",* into *"How the crash of Oct. 27, 1997 could have been predicted"*[38].

The prediction and universality hypothesis can be criticized by physicists, mathematicians and economists, from first principles or on mathematical grounds, not mentioning psychological attitudes. This approach should be taken with caution indeed. The robustness and soundness of models are fundamental questions. I am on this point rather optimistic[42] because physicists know that model hypotheses must be fullfiled before trusting any prediction, but if they are fullfiled, one should not be too cautiously afraid. In fact parameters in the above equations can even, by analogy, receive some reasonable interpretation[41].

Nevertheless the affair was sufficiently publicised in many places, including U. Princeton[42]. It led to many media interviews, invitations and reports[43,44]. There were subsequently interesting discussions with financial managers in Belgium, but their expectations of when a physics model can be applied and how much perfectly reliable theoretical predictions can be, did not allow me to accept their offer to join some Board. Stupid indeed. On the contrary, as a scientist, I only wished to have money put into research and toward undergraduate or graduate student grants. I did not get any support, neither from the Department of Physics nor from the Research Council of the University nor from higher authorities at the federal research level. On the contrary at some time when some promotion of mine should have been considered I was asked to withdraw all papers on econophysics from my publication list and not mention such activities in my CV. No comment!

*Scientific Outreach Time*

On one hand it was necessary to reach co-authors not associated with ULG and to diversify approaches taking into account colleague skills and interests. Other methods than DFA and other empirical laws than LPO had to be investigated, like the so called Order Variability Diagrams[b, 45] and the Zipf Ranking Method[46–49], or the Recurrence Quantification Analysis[50]. Much work was devoted to financial data multifractal aspects[51–53].

Expecting some interest for some simple/classical physics formalism for studying financial systems and formalizing the phenomena, I came up with some model-independent Newtonian description of the financial fluctuations[54,55] and some Boltzmann equation for share price evolution[56].

I remember to have proposed the organization of a conference on application of physics to financial systems and had started to obtain some fund from a local bank and a bank owner having previously been an engineer, Mr. Romagnoli. At some condensed matter expert meeting in Brussels in 1998, I approached the EPS president of the year about organizing such a conference.

He killed my hope of being the first organizer in stating that since he was Irish this should be done in Dublin and he was going to do it. Thus I organized APFA2 in Liège only in 2000. The APFA conferences were organized thereafter in many cities, including Tokyo in 2009.

In between, APFA3 was organized in London. Enticed by such a location, and the timing of the EUR introduction, it was interesting to become the first EUR forfeiter and invent the exchange rates of the EUR prior to its introduction[57]. The time dependent DFA was used and the DFA exponents compared for many currency exchange rates. It was observed that the GBP exchange rate fluctuation correlations were the same as those of the EUR[58]. Having presented the data, I had to answer a question of a Financial Time journalist on "Thus (!), why should we join the EUR?". The journalist was flabergasted when I told him that instead of following the EUR exchange rates, his GBP might lead them. The answer to a question on causes vs. consequences[59] seems still out of reach for many (UK) politicians.

---

[a] As the year progressed, everyday, the week of the crash was more and more becoming precise.

[b] It has been shown that not only long term correlations exist and can be implemented for predictability, but some systematics of the short range correlation functions can also be observed e.g. in foreign currency exchange rate and in futures[45].

### Macro-econophysics

The evolution of econophysics and the lack of support I had (not) could not allow me to remain at a honorable level within such a competitive and fast growing field. Not interested in competing with those who had access to high frequency data, I had to turn toward low frequency data, as in macroeconomy. Since the data is less frequent and not necessarily very reliable the challenge was obvious. Could one apply statistical physics ideas to such macro considerations?

The historical evolution of the world economy was also shocking. Many jobs were lost in Belgium due to delocalization toward Ireland or Poland or further away. Politicians and others were, and are, telling us that "the North" should help "the South". For the best of the western world, was it proclaimed: "World economy globalization is a must". However could one find out whether simple or not physics-like models could come up with some answer on recession durations[60,61] and economic cycles[62]? Psychology is somewhat relevant in such studies: the evolution of the economy was only looked for through some arbitrary criterion on best adapted companies[63], even though delocalization and subsequent increased unemployment in the original company region might have suggested to look also at the not so well adapted companies. Some further work along these lines might be considered as socially relevant as well.

Thus simple modelization was performed on the dynamics of correlations in the evolution of economic entities under varying economic conditions[64–66] remembering that the Berlin wall fall implied drastic economy changes. An attempt to observe economy globalization through the cross correlation distance evolution of GDP's followed[67–70] leading to the surprising fact that the globalization of the economy had somewhat ended and rich countries were recently following various incoherently managed routes in order to recover some controlled (thus "positive") growth[71].

Such globalization studies from a macroeconomic point of view was the source of investigations in a classical Hamiltonian or thermodynamics framework, with some adaptation in terms of network and nanocluster technology so modern nowadays. In so doing, the cluster structure of EU-15 countries was derived from the correlation matrix analysis of several macroeconomic index fluctuations[72,73]. It is remarkable that for the first, and last time, one work was published in an economy journal[74].

In a mechanistic (equilibrium) approach, one can imagine to develop some quantity of interest, some free energy or cost function, as a series in terms of clusters, e.g., ordered according to increasing size, i.e. the number of "spins" in the cluster, as in

$$\exp\left[-\frac{F-F_0}{kT}\right] = -\Sigma_i H_i S_i - \Sigma_{i,j} J_{ij} S_i \times S_j - \Sigma_{i,j,k} K_{ijk} S_i \times S_j \times S_k - \ldots \quad (3)$$

in obvious notations, i.e. each spin $S_i$, located at some site $i$, representing some degree of freedom in an external field $H_i$ and interacting with another through some interaction $J_{ij}$, *etc.*, or similarly

$$\frac{\partial \Phi_i}{\partial t} = A_{ij}\Phi_j + B_{ijk}\Phi_j\Phi_k + C_{ijkl}\Phi_j\Phi_k\Phi_l + \ldots, \quad (4)$$

For some information flux $\Phi_i$. A vector generalization is immediately thought of[75]. A Langevin equation might next be written.

Therefore it is emphasized that one should describe financial markets or an economy through degrees of freedom which may be both qualitative and quantitative in nature, – each node being the siege of a quite complicated mathematical entity, surely not a scalar. Interactions between the various degrees of freedom of different nodes can occur, through a matrix form; each link having a weight and a direction. That interaction matrix might not be symmetric. The more so if time delays are taken into account.

Along the above lines one should insist on time delay effect in the information flow. It is often said that some agent is leading or following the evolution of another, – even though only equal time correlations are investigated. Except if time scales are quite long, the conclusion does not seem correct. In fact it has been shown through studies of macroeconomic indices and simulations that the influence of information flow in the formation of economic cycles is highly relevant[68] and delayed information flow does markedly affect economy system evolutions[76,77]. Although this has not been investigated along the discussed lines such a remark should carry true into microeconomy.

### Conclusion

There is much data available in financing, banking, option markets, stocks and currency exchange rates, discount and interest rates[78]. When such data must be bought the cost is beyond the reach of small and poor research groups. Fortunately, when there is no money, one has to rely on creativity and implement original ideas. One is immediately tempted to play statistics, hopefully expecting some understanding of financial matters, on many levels of observation: individual income, individual expenses, checking accounts and savings, public or private accounts, volumes, debts and credits, tellers, dealers, bank outlets, businesses, governments, ... turning toward micro- or macro-economy considerations.

Several scientific issues can be found e.g. (i) creating models based on financial insights and mathematical principles, (ii) calibrating models based on market information, and (iii) simulating models using specific algorithms. Even modern, like non-extensive, statistical physics might profit of finance/economy measures[79]. In Liège, Belgium several topics have been touched upon, due to intellectual default and scientific necessity. Two types of themes, due to data availability, have been studied: stock markets and foreign currency exchange rates; there are based on different goals and strategies of "players" (in other words "traders"). Based on, right or wrong analogies, cooperative effects like those seen at phase transitions may appear as an open door. This is a gateway for physicists to complement or surpass mathematicians who have invaded economy faculty or arbitration rooms or brokerage firms in the past. A defect might be that physicists know the limits of model understanding, and do honestly question their findings. Many economists have theories based on mathematics and statistics, - which often tell everything you want to say or to hear. Alas it was found through the Belgian environment that financial matters, through a physicist looking glass, are not yet a meeting place for those who have open minds.

Economists scorn physics models because the former ones claim to have also endeavoured the prediction of short term fluctuations by examining past fluctuations, as in chart analysis, or approximating stock prices by wave structures. These models can be a source of profits when they turn out to be a self-fulfilling prophecy. This can happen if many market actors believe in such theories. Therefore economists are sending warnings about their own techniques, though not seeing differences with respect to physicist approaches.

In concluding, one could regret that following a good start, the econophysics community in Belgium is so small, and might be disappearing. Attempts toward spin-offs, or spin-outs, existed. They lasted the time of a rose. I can be pleased to have been called for organizing curricula elsewhere[c], to have been invited in Ph. D. jury or at international conferences. I believe that many important results have been obtained through my catalytic effect. Spectacular results are necessary for maintaining enthusiasm. However media impacts often lead to grief, jealousy, contempt.

Due to the small size of the country and budgets one cannot expect that a young research field is appreciated nor can be sustained in every university. It was great that one university professor in the U. of Liège Faculty of Economy had an open mind. Others did/do not have so. Lectures were proposed to the Faculty of Sciences in Liège for inclusion into the physics curriculum, but they only survived a couple of years. Students were told that this is not physics. Nevertheless overviews of econophysics ideas are still included in introductory courses in statistical physics.

More impact on academia and curricula can be forecasted if there is some practical effect, - if jobs are open to econophysics students per se. Very strangely one member of the SUPRATECS group who went into banking is a chemist who never worked on econophysics.

*Acknowledgements*

---

[c] Thanks to those who asked me for doing so. I feel very honored